\documentclass[conference,10pt,letterpaper]{IEEEtran}
\usepackage[T1]{fontenc}
\usepackage{cite}
\usepackage{amsmath,amssymb}
\usepackage{graphicx}
\usepackage{xcolor}
\usepackage{overpic}
\usepackage{booktabs}
\usepackage[checkfloat,nospread]{flushend}
\usepackage[hidelinks]{hyperref}
\newcommand{\EbNo}{\ensuremath{E_b/N_0}}
\newcommand{\RR}{\mathbb{R}}

\newcommand{\photolabel}[1]{%
 \makebox(0,0){\setlength{\fboxsep}{1pt}%
 \colorbox{white}{\sffamily\bfseries\fontsize{6.5}{7.5}\selectfont #1}}}

\begin{document}
\title{Design and Over-the-Air Evaluation of a\\Digital Twin-Driven Site-Specific Neural Receiver\\for 5G NR on NVIDIA DGX Spark}
\hypersetup{pdftitle={Design and Over-the-Air Evaluation of a Digital Twin-Driven Site-Specific Neural Receiver for 5G NR on NVIDIA DGX Spark}}
\author{
    \IEEEauthorblockN{Qiuyuan Chen, Chengyang Yao, Liuchang Zhuo, Yuquan Sun, Cunhua Pan\textsuperscript{\textdagger}, Hong Ren, Jiangzhou Wang}
    \IEEEauthorblockA{National Mobile Communications Research Laboratory, Southeast University, Nanjing, China \\
    Email: njcqy2021@qq.com, \{220260879, 230248854, yq.sun, cpan, hren, j.wang\}@seu.edu.cn}
}
\maketitle

\begin{abstract}
We design and evaluate a site-specific neural receiver for the 5G New Radio (NR) uplink. A corridor digital twin built from a radar scan and refined manually provides 28,671 ray-traced channel records for training an eight-round convolutional graph neural network with pilot-aware inputs and auxiliary channel supervision. The GPU on NVIDIA DGX Spark evaluates time--frequency convolutions in parallel across the received grid to accelerate inference. The receiver uses the standard NR uplink waveform and reference signals and feeds soft bits to the existing transport-block decoder. In over-the-air tests with NI USRP B210 radios under common radio settings, the site-trained receiver achieves the highest mean uplink rate at all five positions. At a non-line-of-sight corner, it reaches 16.2~Mb/s, compared with 4.1~Mb/s for the conventional receiver and 2.4~Mb/s for the 3GPP-trained neural receiver. These results demonstrate the feasibility of integrating neural reception into the 5G NR physical layer and show its potential for AI-RAN deployments.
\end{abstract}


\section{Introduction}
Neural receivers jointly learn channel processing and soft-bit prediction while retaining the cellular waveform, pilots, and channel code. The training channels determine which propagation conditions the receiver learns from. A digital twin of the deployment site can supply channels that reflect the local geometry and radio materials. This study evaluates whether training on these channels improves measured 5G NR uplink rates at line-of-sight (LoS) and non-line-of-sight (NLoS) positions.

Cammerer \emph{et al.} introduced time--frequency convolution and user-wise graph aggregation for 5G NR reception~\cite{cammerer2023}. Wiesmayr \emph{et al.} used ray-traced channels for site adaptation and evaluated simulated block error rates and GPU inference timing~\cite{wiesmayr2024}. Measurement-based studies have examined site-specific finetuning on captured uplink signals~\cite{baytekin2026} and cross-site transfer~\cite{luostari2026}. This work builds on the existing neural receiver architecture and software interfaces~\cite{neuralrxsoftware} and measures uplink rates in the same corridor used to build the digital twin.

A radar scan provides the initial 3D geometry of the test corridor. Manual refinement and material assignment prepare the mesh for Sionna RT~\cite{sionnart2023}, which generates channel impulse responses (CIRs). These CIRs drive NR link simulations in Sionna~\cite{sionna2022} to train the detector for deployment on a GPU and software-defined radio testbed.

The detector runs in the base station's 5G NR physical uplink shared channel (PUSCH) chain. It uses the received orthogonal frequency-division multiplexing (OFDM) grid and features from standard demodulation reference signals (DMRSs) to produce soft bits for the existing transport-block decoder. This integration lets a base station deploy site-specific detection with existing NR transmitters. The main contributions are as follows:

\begin{enumerate}
\item \emph{Digital twin and training data.} The material-labeled corridor digital twin supplies ray-traced CIRs for training and evaluation.
\item \emph{Neural receiver training and deployment.} The eight-round detector uses pilot features and auxiliary channel supervision. Parallel GPU execution across resource elements accelerates soft-bit inference on DGX Spark.
\item \emph{Over-the-air validation.} USRP B210 tests under common radio settings compare three receivers. The site-trained receiver leads at all five positions, with the largest absolute gain over LS/LMMSE at an NLoS corner.
\end{enumerate}

\section{System Model and Scene-Conditioned Data}
\subsection{NR Uplink Model}
\begin{figure}[!t]
\centering
\includegraphics[width=\columnwidth]{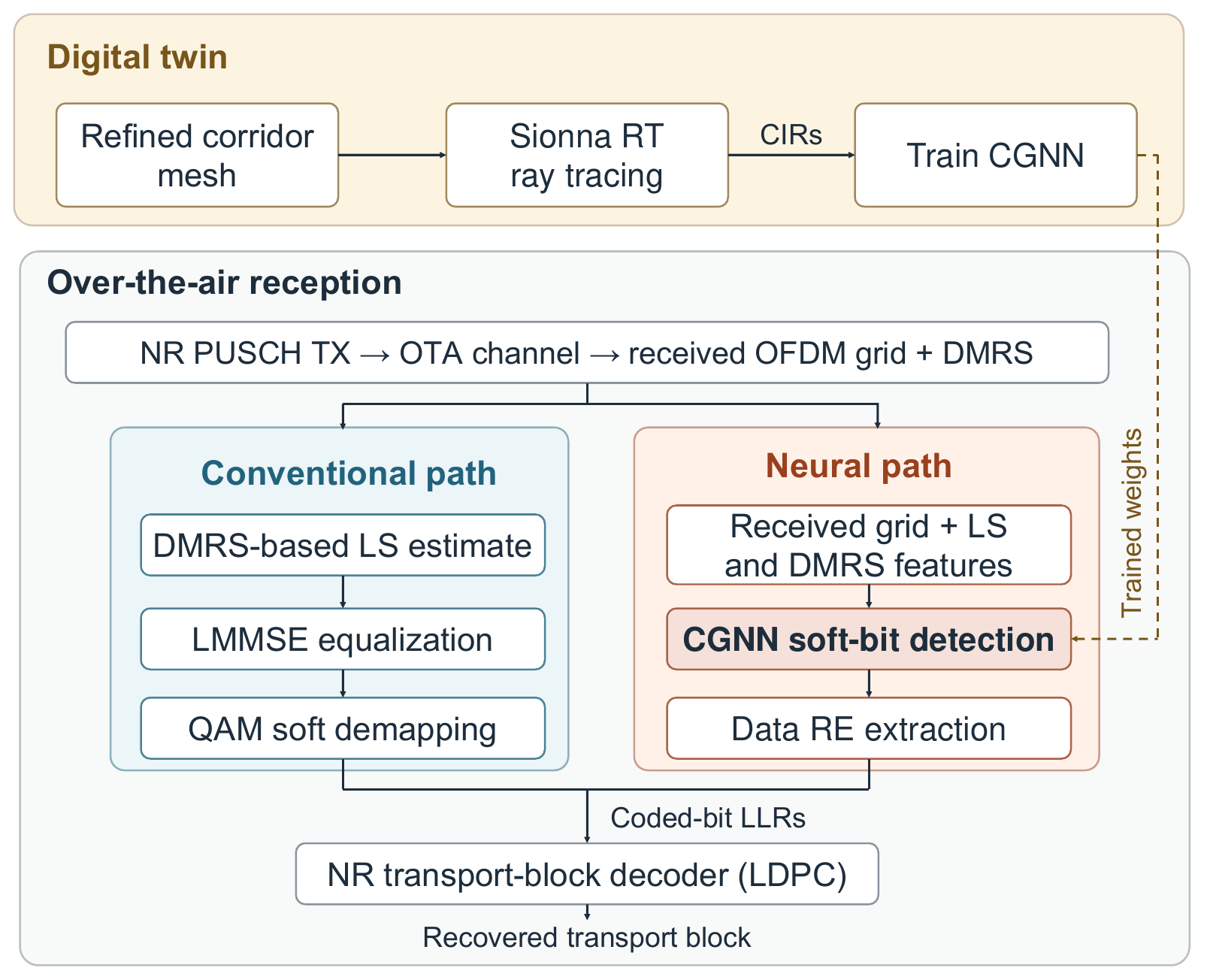}
\caption{Offline training from ray-traced channels and the two receiver paths for over-the-air 5G NR PUSCH reception.}
\label{fig:receiverpaths}
\end{figure}

Fig.~\ref{fig:receiverpaths} separates offline training with ray-traced CIRs from over-the-air PUSCH reception. The conventional receiver uses DMRSs for least-squares (LS) channel estimation, followed by linear minimum mean-square error (LMMSE) equalization and QAM soft demapping. The convolutional graph neural network (CGNN) uses the received grid, pilot-domain LS estimates, and DMRS positions to generate coded-bit log-likelihood ratios (LLRs). Both paths feed the same NR transport-block decoder. Section~\ref{sec:neural} details the neural architecture.

We consider a single user transmitting one spatial layer to a single-antenna receiver. Under the frequency-domain OFDM channel model, the received resource element is
\begin{equation}
 Y_{k,t}=H_{k,t}X_{k,t}+W_{k,t}, \qquad
 W_{k,t}\sim\mathcal{CN}(0,N_0),
 \label{eq:link}
\end{equation}
where $k$ and $t$ index subcarriers and OFDM symbols, respectively. The simulator uses the NR PUSCH transmitter and transport-block decoder.

Table~\ref{tab:system} lists the shared waveform configuration. MCS table~1, index~14 corresponds to modulation order $Q_m=4$ and target code rate $553/1024$~\cite{3gpp38214}. Thus, the experiment uses 16-QAM. The effective transport-block rate also includes NR size selection and overhead. The energy axis is information-bit \EbNo, converted to noise variance by the simulator using the modulation order, code rate, and resource-grid overhead.

\begin{table}[t]
\caption{Radio Settings and Proposed Neural Receiver}
\label{tab:system}
\centering
\begin{tabular}{@{}ll@{}}
\toprule
Parameter & Value\\
\midrule
Carrier frequency & 3.30948 GHz\\
Users / transmit / receive antennas & 1 / 1 / 1\\
Subcarrier spacing & 30 kHz\\
Allocated resource blocks / subcarriers & 51 / 612\\
OFDM symbols / cyclic prefix & 14 / normal\\
MCS table / index & 1 / 14\\
Modulation / target code rate & 16-QAM / $553/1024$\\
DMRS mapping / configuration type & A / 1\\
DMRS length / additional position & 1 / 1\\
Neural iterations / state dimension & 8 / 56\\
Stored model scalars (both heads) & 434,948\\
Simulation decoder BP iterations & 20\\
Simulation \EbNo & $-10$ to $9$ dB, step 1 dB\\
\bottomrule
\end{tabular}
\end{table}

\subsection{Corridor Digital Twin and Channel Generation}
\begin{figure}[!t]
\centering
\includegraphics[width=0.8\columnwidth]{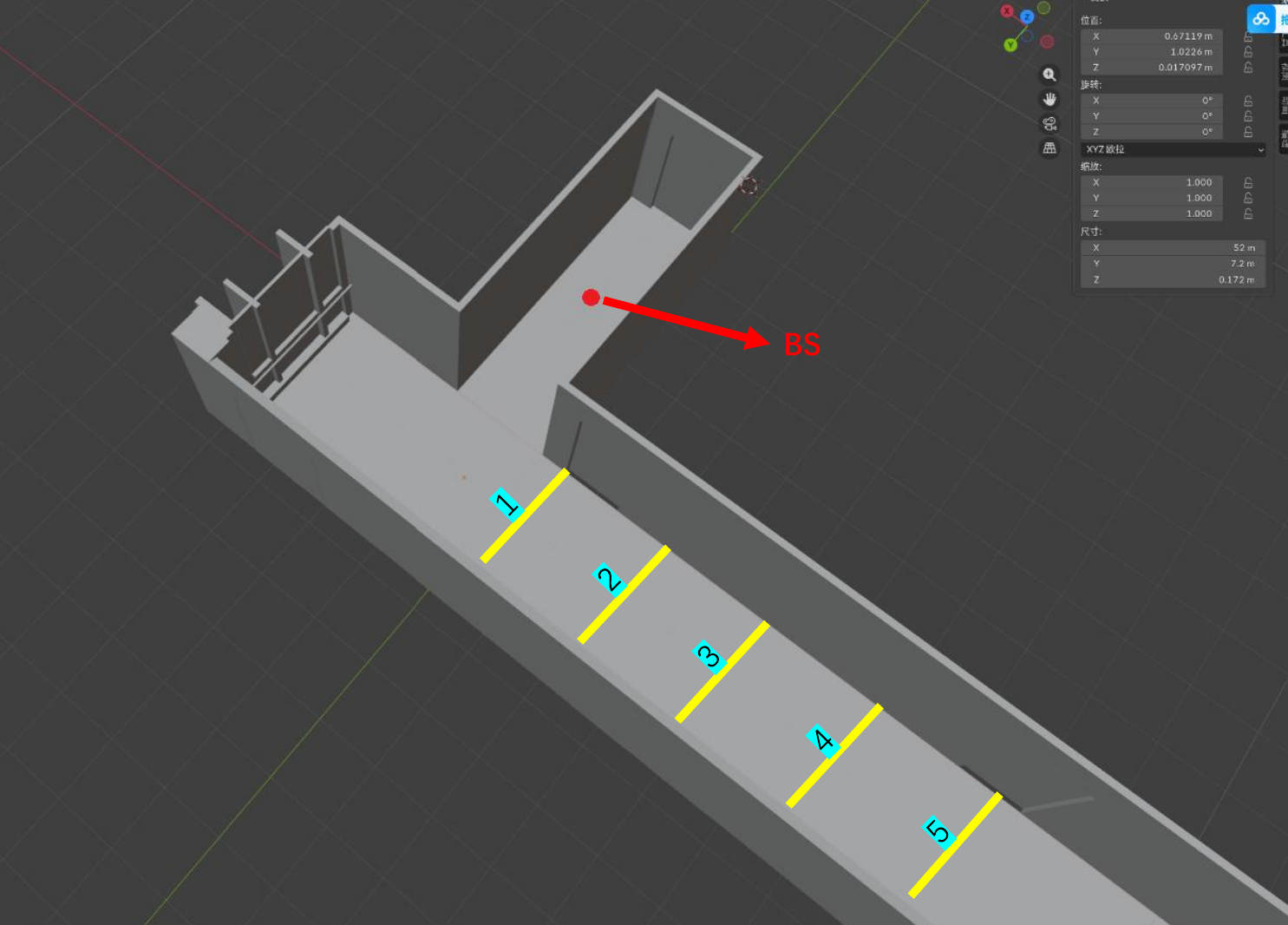}\\
{\footnotesize (a)}\\[4pt]
\includegraphics[width=\columnwidth]{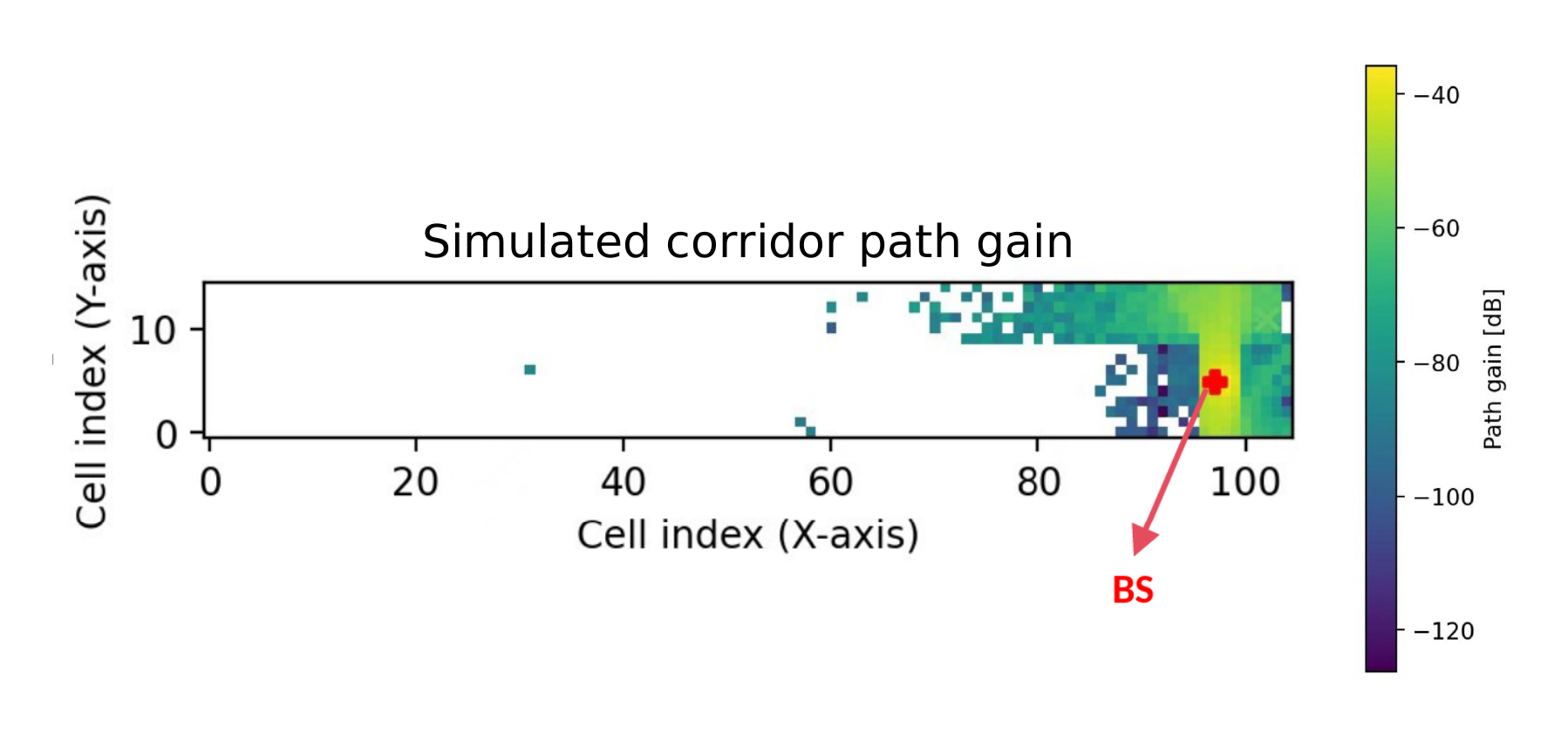}\\
{\footnotesize (b)}
\caption{Corridor digital twin. (a) Radar-derived mesh refined in Blender. Yellow lines mark locations 1--5. (b) Sionna RT path-gain map. Red markers indicate the fixed base station.}
\label{fig:sceneplan}
\end{figure}

The digital twin starts from a radar scan of the physical corridor shown in Fig.~\ref{fig:testbed}. We reconstruct a 3D mesh from the scan and manually refine the corridor geometry in Blender, as shown in Fig.~\ref{fig:sceneplan}(a). The mesh has eight components representing the floor, walls, ceiling, doors, and windows, with assignments for concrete, ceiling board, wood, glass, and metal. Sionna RT uses the refined mesh and its radio material assignments to simulate propagation. Fig.~\ref{fig:sceneplan}(b) shows the resulting path-gain map and marks the fixed base station. For uplink channel generation, user transmitters are sampled within the modeled corridor with heights between 0 and 2~m and configured horizontal speeds between 0 and 1~m/s.

For a simulated link, the multipath representation is
\begin{equation}
 \begin{aligned}
 h(\tau,t)&=\sum_{p=1}^{P}a_p(t)\delta(\tau-\tau_p),\\
 H_{k,t}&=\sum_{p=1}^{P}a_p(t)e^{-j2\pi f_k\tau_p},
 \end{aligned}
 \label{eq:cir}
\end{equation}
where $a_p(t)$ and $\tau_p$ denote a path coefficient and delay. Channel generation uses a maximum interaction depth of five, with LoS, specular reflection, diffuse reflection, and diffraction enabled. The geometry and material assignments thereby determine the multipath population exposed to the receiver during training.

The simulator normalizes each channel realization to unit average energy before adding noise, so its error-rate curves characterize reception over normalized channel structure. Section~\ref{sec:ota} reports the uplink rates measured over the air.

\subsection{Dataset Conversion and Split}
The raw archive contains 32,768 samples in 256 batches. Removing 996 samples without valid paths leaves 28,671 training and 3,101 evaluation records. Conversion assigns each retained sample independently to evaluation with probability 0.1, forming a random holdout within the corridor scene.

Each record stores complex coefficients and real delays for 128 path entries and 14 time samples. Paths are ranked by maximum coefficient magnitude across antennas and time, truncated, and zero-padded as needed. The stored time samples provide the simulated slot's channel sequence.

Training and evaluation resample the respective channel records with fresh payloads and noise, producing many transmissions over the finite scene-conditioned channel population.

\begin{figure*}[!t]
\centering
\includegraphics[width=\textwidth]{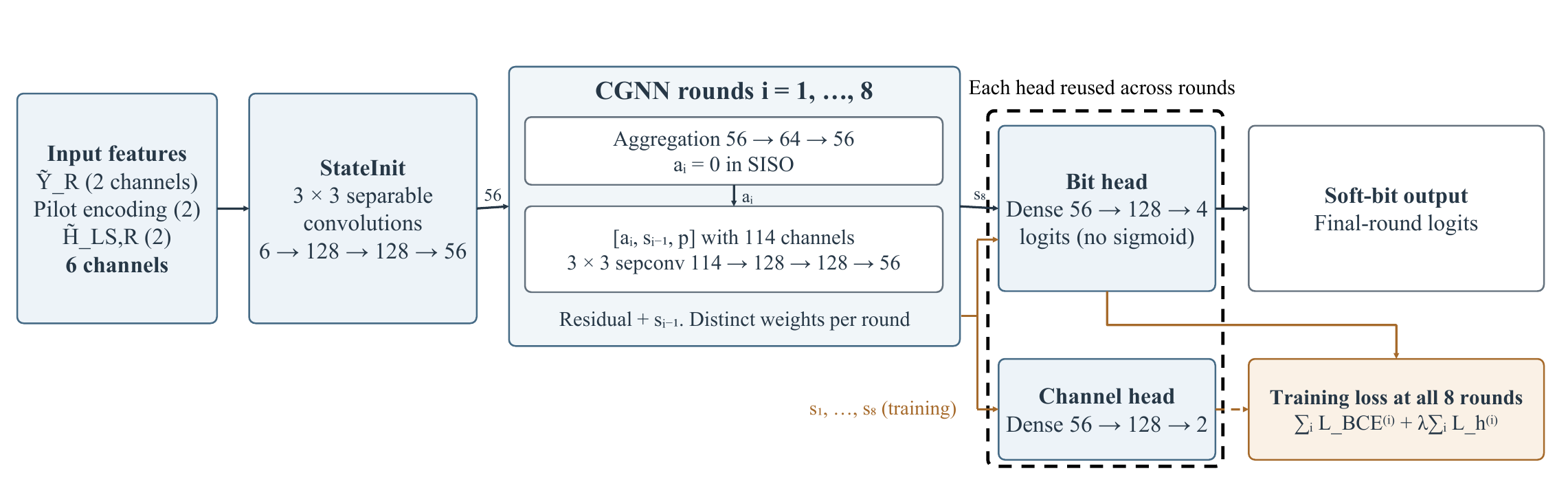}
\caption{SISO CGNN with pilot-aware state initialization, eight residual update rounds, and two readout heads shared across rounds. The final-round bit logits feed the NR decoder, and the channel readouts enter the training loss.}
\label{fig:neural}
\end{figure*}

\section{Neural Receiver Architecture and Training}
\label{sec:neural}
We configure the established CGNN architecture~\cite{cammerer2023,wiesmayr2024,neuralrxsoftware} for one user, one receive antenna, and one MCS. Fig.~\ref{fig:neural} shows how the receiver initializes the state from received samples and pilot features, refines it over eight rounds, and applies the bit and auxiliary channel heads.

\subsection{Scene-Conditioned Learning}
Let $q_{\mathrm{scene}}(H)$ be the empirical distribution of normalized frequency-domain channels obtained from the training CIR records. At training stage $s$, the objective is
\begin{equation}
 \mathcal J_s(\theta)=\mathbb E_{\substack{H\sim q_{\mathrm{scene}},\,\mathbf c,\,W\\
 \gamma\sim\mathcal U[a_s,b_s]}}
 \left[\mathcal L_{\mathrm{bit}}(\theta)+\lambda_s\mathcal L_h(\theta)\right],
 \label{eq:scene_objective}
\end{equation}
where $\theta$ collects the network weights, $\mathbf c$ denotes freshly generated coded bits, $\gamma$ is \EbNo\ in dB, and $W$ is AWGN with variance determined by $\gamma$. Adam successively optimizes these stage objectives. Geometry and material information affect $q_{\mathrm{scene}}$ and hence the learned weights; the ray tracer also supplies the reference $H$ for auxiliary supervision. Deployment uses observed samples and DMRS, with scene information carried by the trained weights. Reference channels and scene geometry are used offline.

\subsection{Pilot-Aware Input Features}
Let $B,U,K,T,N_r$ denote batch size, users, subcarriers, OFDM symbols, and receive antennas. The state layout is $[B,U,K,T,C]$, with feature width $C$. At pilot locations $\mathcal P_u$, raw single-port estimates are $Y_{k,t}/X_{k,t}^{\mathrm{DMRS}}$. The PUSCH estimator averages neighboring pilot pairs in frequency for code-division-multiplexing handling, then applies nearest-neighbor interpolation to the full grid. This supplies a channel prior while the received samples retain information lost by averaging and interpolation.

Two positional features identify distance to the user's pilots along the time and frequency axes:
\begin{equation}
 d_u^t(t)=\min_{(k_p,t_p)\in\mathcal P_u}|t-t_p|,\quad
 d_u^f(k)=\min_{(k_p,t_p)\in\mathcal P_u}|k-k_p|.
 \label{eq:position}
\end{equation}
The distances are standardized separately along their respective axes, with zero-variance features kept at zero, then broadcast across the other axis to form $\mathbf p_u\in\RR^{K\times T\times2}$. These are distances to pilot indices, rather than physical coordinates or measured channel quality. They allow the same convolutional kernels to condition processing on pilot proximity.

Received samples and LS estimates are split into real and imaginary parts. Per example, the receiver applies
\begin{equation}
 \begin{aligned}
 \alpha&=\left(\frac{\|\mathbf y_R\|_F^2}{2N_rKT}\right)^{-1/2},\\
 \widetilde{\mathbf y}_R&=\alpha\mathbf y_R,\qquad
 \widetilde{\mathbf h}_{\mathrm{LS},R}=\alpha\mathbf h_{\mathrm{LS},R}.
 \end{aligned}
 \label{eq:normalization}
\end{equation}
The feature tensor concatenates $[\widetilde{\mathbf y}_R,\mathbf p_u,\widetilde{\mathbf h}_{\mathrm{LS},R}]$, giving $4N_r+2$ channels. Thus, for this experiment the input is $[B,1,612,14,6]$: two received-signal, two position, and two channel-estimate features. No explicit noise variance is appended. Equation~\eqref{eq:normalization} sets real-component mean square to one; it is distinct from the propagation-channel normalization in Section~II.

\subsection{Separable Convolutions and State Initialization}
The initialization and update blocks use depthwise-separable $3\times3$ convolutions. A depthwise filter first processes each feature channel over neighboring resource elements; a $1\times1$ pointwise filter then mixes channels. With depth multiplier one, input/output widths $C_{\mathrm{in}},C_{\mathrm{out}}$, and bias, the parameter count is
\begin{equation}
 P_{\mathrm{sep}}=9C_{\mathrm{in}}+C_{\mathrm{in}}C_{\mathrm{out}}+C_{\mathrm{out}}.
 \label{eq:sep}
\end{equation}
This factorization reduces weights relative to a full $3\times3$ convolution while allowing time--frequency interaction. All convolutions have stride one and same padding; there is no pooling or downsampling. The initializer has widths $6\rightarrow128\rightarrow128\rightarrow56$, ReLU after the first two layers, and a linear final layer. It produces $\mathbf S^{(0)}\in\RR^{B\times1\times612\times14\times56}$. The 56 features are learned latent states, rather than 56 physical channel taps.

\begin{table}[t]
\caption{Neural Modules and Stored Parameter Counts}
\label{tab:neural}
\centering
\begin{tabular}{@{}llr@{}}
\toprule
Module & Feature widths & Parameters\\
\midrule
Initializer & $6\!\to\!128\!\to\!128\!\to\!56$ & 26,990\\
Aggregation / round & $56\!\to\!64\!\to\!56$ & 7,288\\
State update / round & $114\!\to\!128\!\to\!128\!\to\!56$ & 41,786\\
Bit head (shared) & $56\!\to\!128\!\to\!4$ & 7,812\\
Channel head (shared) & $56\!\to\!128\!\to\!2$ & 7,554\\
\midrule
Total (8 rounds) & & 434,948\\
\bottomrule
\end{tabular}
\end{table}

\subsection{User Aggregation and Residual State Updates}
Each round $i=1,\ldots,8$ first applies a pointwise aggregation MLP $\phi_i$, with widths $56\rightarrow64\rightarrow56$, to each user's previous state. The hidden layer uses ReLU and the output is linear. For an active user $u$, aggregation averages the transformed states of the other active users:
\begin{equation}
 \mathbf A_u^{(i)}=
 \frac{\sum_{v\ne u}m_v\phi_i(\mathbf S_v^{(i-1)})}
 {\max(1,\sum_v m_v-1)},
 \label{eq:aggregation}
\end{equation}
where $m_v$ is the active-user mask. Aggregation is performed independently at each resource element. Here $U=1$, so $\mathbf A_u^{(i)}=0$ and the active processing is iterative time--frequency convolution.

The update concatenates the aggregate, previous state, and pilot encoding, yielding $56+56+2=114$ channels. Its residual mapping is
\begin{equation}
 \mathbf S_u^{(i)}=\mathbf S_u^{(i-1)}+
 F_i\!\left([\mathbf A_u^{(i)},\mathbf S_u^{(i-1)},\mathbf p_u]\right).
 \label{eq:update}
\end{equation}
The separable-convolution widths are $114\rightarrow128\rightarrow128\rightarrow56$, with two ReLUs and a linear output. No activation follows the residual addition. This preserves a direct state path while learning a correction. The eight $F_i$ and $\phi_i$ have independent weights; within a round, weights are reused across users and resource elements. Repeated local updates expand the convolutional context while preserving grid resolution. Table~\ref{tab:neural} gives the stored counts, including aggregation modules and the auxiliary head.

\subsection{Soft-Bit and Auxiliary Channel Readouts}
Two pointwise MLPs read each updated state. Each has a 128-unit ReLU hidden layer and a linear output, with widths $56\rightarrow128\rightarrow4$ for bits and $56\rightarrow128\rightarrow2$ for the channel. The two heads have separate weights, but each head is shared across all eight rounds. The bit tensor has shape $[B,1,612,14,4]$, one logit per 16-QAM bit. There is no output softmax or hard decision. The channel tensor has shape $[B,1,612,14,2]$, representing real and imaginary channel components.

Resource-grid demapping removes pilot/non-data positions and layer demapping restores coded-bit order. At inference, the final-round logits are passed to the standard NR transport-block decoder, including LDPC decoding; the decoder is outside the neural network. The auxiliary channel output is supervised during S3 training, but is not fed into a separate equalizer or back into the next iteration. Consequently, the learned receiver maps pilot-aided observations directly to soft bits, rather than implementing a cascade of a learned channel estimator and a conventional equalizer.

\subsection{Multi-Iteration Supervision}
Training uses regenerated \emph{coded} bits, not decoded payload decisions. Here, 7,956 data resource elements give $N_c=31,824$ coded-bit targets per slot. Let $c_{b,j}\in\{0,1\}$ be a target and $\ell_{b,j}^{(i)}$ its logit. The bit objective is
\begin{equation}
 \mathcal L_{\mathrm{bit}}=\sum_{i=1}^{8}\frac{1}{BN_c}
 \sum_{b,j}\left[\log(1+e^{\ell_{b,j}^{(i)}})-c_{b,j}\ell_{b,j}^{(i)}\right],
 \label{eq:bce}
\end{equation}
implemented by a numerically stable BCE-from-logits operation. A positive logit favors bit one under this convention. Auxiliary supervision compares each readout against the simulator's channel label:
\begin{equation}
 \begin{aligned}
 \mathcal L_h&=\sum_{i=1}^{8}\frac{1}{2N_rBKT}
 \sum_b\big\|\mathbf H_{R,b}-\widehat{\mathbf H}_{R,b}^{(i)}\big\|_F^2,\\
 \mathcal L&=\mathcal L_{\mathrm{bit}}+\lambda\mathcal L_h.
 \end{aligned}
 \label{eq:chest}
\end{equation}
The channel target is the normalized propagation channel in real form, not the noisy LS input multiplied by $\alpha$. Both losses sum over rounds without a factor $1/8$, while averaging over examples and supervised coordinates within each round. Intermediate supervision supplies gradients to earlier states as well as the final state. Training backpropagates through the neural processing up to its soft-bit and channel readouts; the transport-block decoder is used for evaluation and reception.

\begin{figure}[!t]
\centering
\includegraphics[width=\columnwidth]{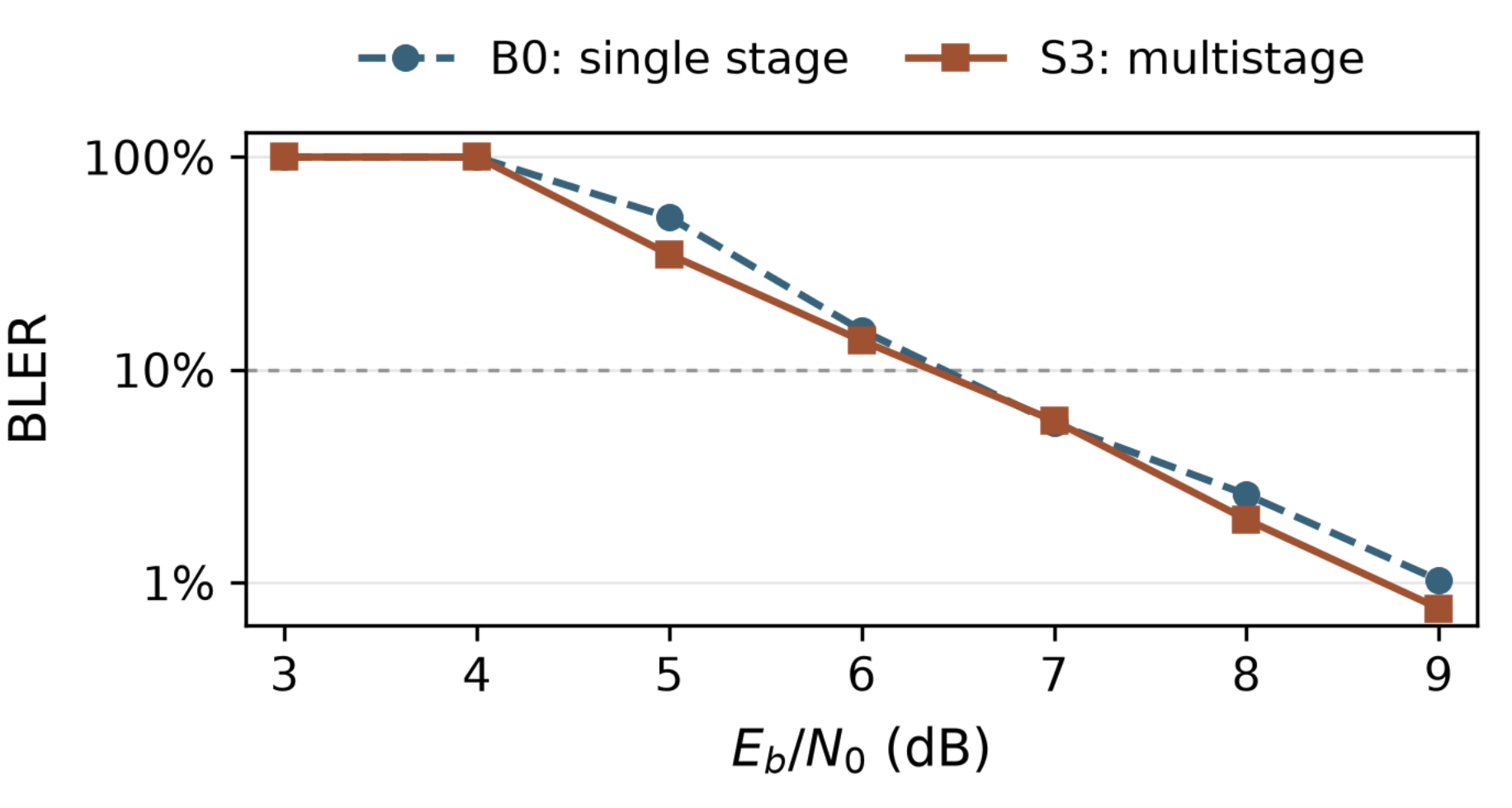}
\caption{Simulated transport-block BLER near the waterfall for B0 and S3.}
\label{fig:errors}
\end{figure}

\begin{figure*}[!t]
\centering
\begin{minipage}[t]{0.405\textwidth}
\vspace{0pt}\centering
\begin{overpic}[percent,width=\linewidth]{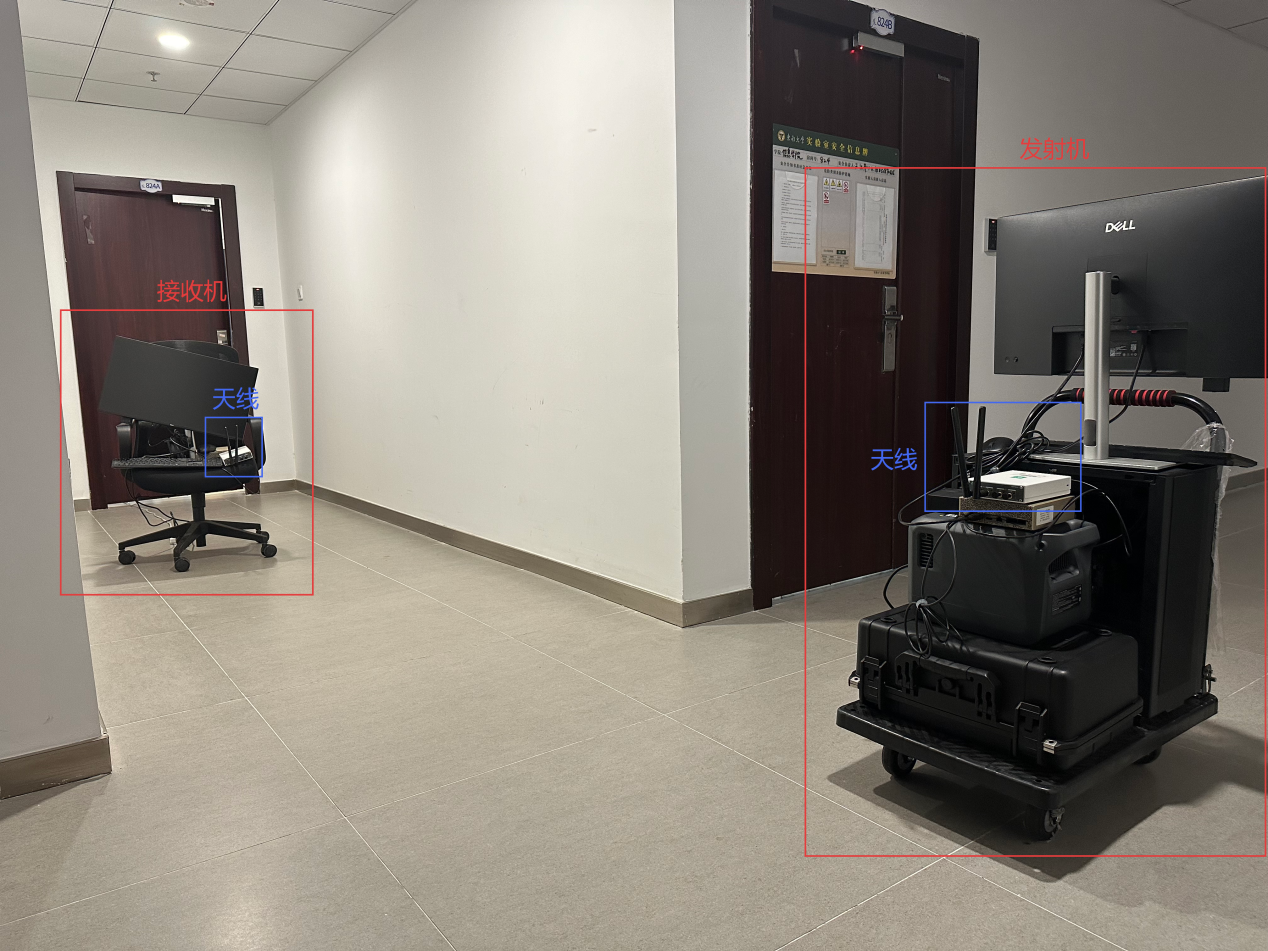}
 \put(15.2,52.0){\photolabel{Fixed BS (RX)}}
 \put(83.1,63.2){\photolabel{Mobile UE (TX)}}
 \put(18.5,43.6){\photolabel{Antenna}}
 \put(70.5,38.8){\photolabel{Antenna}}
\end{overpic}\\[2pt]
\footnotesize (a) Fixed BS and mobile UE in the test corridor.
\end{minipage}\hfill
\begin{minipage}[t]{0.235\textwidth}
\vspace{0pt}\centering
\begin{overpic}[percent,width=\linewidth]{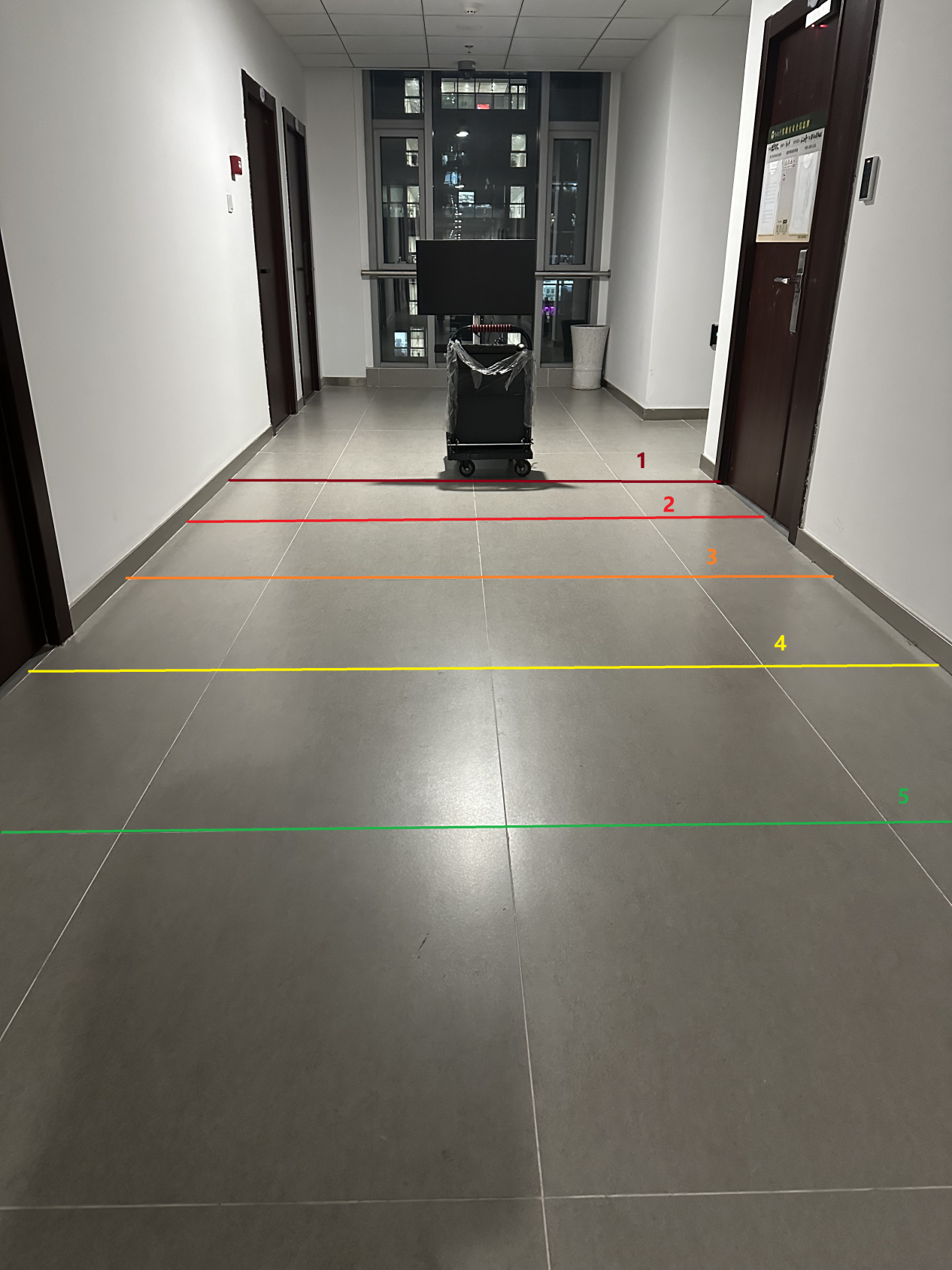}
 \put(50.5,63.8){\photolabel{1}}
 \put(52.7,60.3){\photolabel{2}}
 \put(56.1,56.2){\photolabel{3}}
 \put(61.5,49.4){\photolabel{4}}
 \put(71.2,37.3){\photolabel{5}}
\end{overpic}\\[2pt]
\footnotesize (b) UE test positions 1--5.
\end{minipage}\hfill
\begin{minipage}[t]{0.28\textwidth}
\vspace{0pt}\centering
\includegraphics[width=\linewidth]{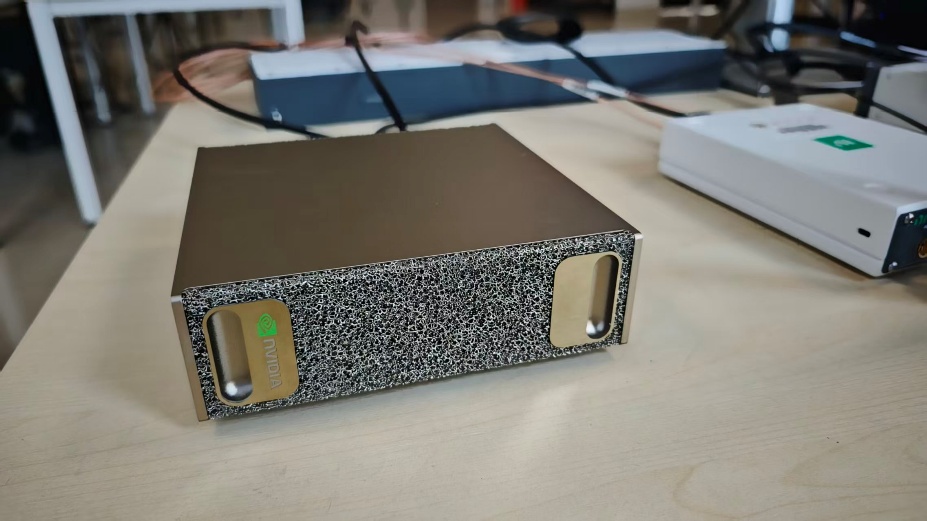}\\[2pt]
\footnotesize (c) NVIDIA DGX Spark.\\[5pt]
\includegraphics[width=\linewidth]{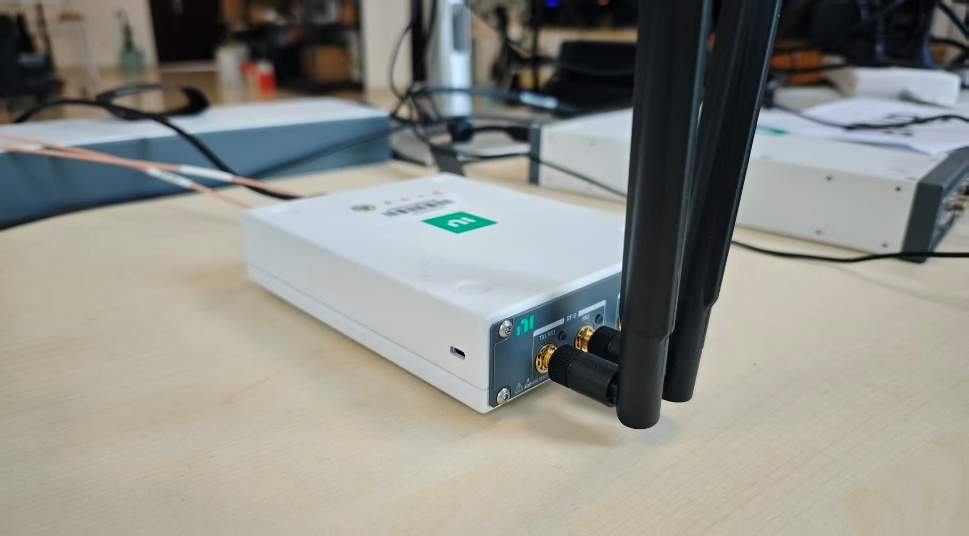}\\[2pt]
\footnotesize (d) NI USRP B210.
\end{minipage}
\caption{Hardware testbed and measurement locations. The fixed base station (BS) and movable user equipment (UE) each use a DGX Spark and a USRP B210. Position~1 is LoS and positions~2--5 are NLoS.}
\label{fig:testbed}
\end{figure*}

\subsection{Training Configurations and Results}
Table~\ref{tab:training} compares single-stage training with bit loss only (B0, 10,000 updates) and three-stage training with an auxiliary channel loss (S3, 70,000 updates). Both use Adam, a batch size of 32, and bit supervision at all eight receiver rounds. Each batch resamples stored channels and payloads, generates fresh AWGN, and draws nominal \EbNo\ uniformly from the stage's dB interval.

\begin{table}[t]
\caption{Training Settings; Intervals Refer to \EbNo\ in dB}
\label{tab:training}
\centering
\begin{tabular}{@{}lrrrr@{}}
\toprule
Run / stage & Updates & Learning rate & Interval & $\lambda$\\
\midrule
B0 & 10,000 & $10^{-3}$ & $[-10,10]$ & 0\\
S3 / 1 & 10,000 & $10^{-3}$ & $[-10,10]$ & 0.02\\
S3 / 2 & 30,000 & $10^{-3}$ & $[-10,5]$ & 0.02\\
S3 / 3 & 30,000 & $5\!\times\!10^{-4}$ & $[-5,5]$ & 0.01\\
\bottomrule
\end{tabular}
\end{table}

S3 narrows the sampling distribution and concentrates its final stage on $[-5,5]$~dB. B0 and S3 are historical training configurations with different total updates and loss settings; their comparison characterizes the combined schedule and supervision choice.

We evaluate B0 and S3 on the 3,101-record indoor channel holdout using the same NR transport-block decoder with up to 20 belief-propagation iterations. Fig.~\ref{fig:errors} shows the transport-block error rate (BLER) near the waterfall, which demonstrates that the three stage training method is effective. 

\begin{figure}[!t]
\centering
\includegraphics[width=\columnwidth]{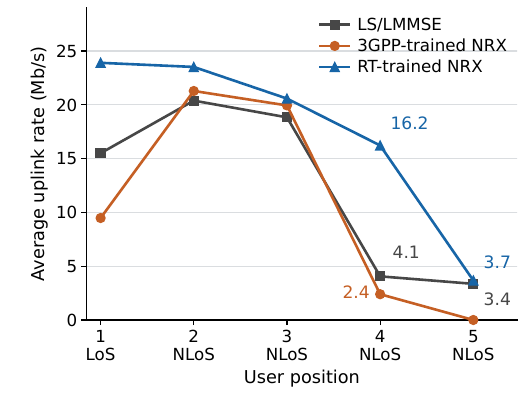}
\caption{Mean measured uplink rate under common radio settings, RF gains, and test procedure. Position~1 is LoS and positions~2--5 are NLoS.}
\label{fig:ota}
\end{figure}

\section{Hardware Deployment and Over-the-Air Validation}
\label{sec:ota}
\subsection{Test Environment, User Positions, and Platform}
We deployed and tested the receiver in an indoor corner corridor (Fig.~\ref{fig:testbed}). A fixed base station receives from a movable user terminal at one line-of-sight (LoS) position (1) and four non-line-of-sight (NLoS) positions (2--5).

Both the base-station side and user side use an NVIDIA DGX Spark and an NI USRP B210. NVIDIA's Sionna Research Kit provides a reference platform for integrating neural receivers into 5G NR hardware testbeds~\cite{sionna-rk}. The three compared receivers use the same carrier, PRB allocation, MCS, SISO configuration, transmit-gain settings, and test procedure. Table~\ref{tab:system} lists the waveform settings, including 51~PRBs, 30~kHz subcarrier spacing, and MCS table~1/index~14 (16-QAM).

\subsection{Receiver Integration and Compared Methods}
We deploy the neural path in Fig.~\ref{fig:receiverpaths} using the existing framework~\cite{neuralrxsoftware}. The GPU executes the convolutions and pointwise readouts in parallel across the $612\times14$ resource grid within each slot. Export permutes the core bit readout from $[B,U,K,T,Q_m]$ to the Aerial layout $[B,Q_m,U,K,T]$ and reverses the logit polarity. The downstream adapter follows this convention when extracting data resource elements and restoring coded-bit order for LDPC decoding.

The three field methods are \emph{LS/LMMSE}, using LS channel estimation and conventional LMMSE reception; \emph{3GPP-trained NRX}, the original neural receiver trained using 3GPP standard channel models; and \emph{RT-trained NRX}, the proposed scene-conditioned receiver. The RT-trained NRX deployed in the field tests was trained using the S3 configuration in Table~\ref{tab:training}. All three receivers share the radio configuration, gains, and test procedure described above.

\subsection{Measured Uplink Rates}
Fig.~\ref{fig:ota} compares the mean measured uplink rates at the five physical positions. RT-trained NRX leads both baselines at all five positions, achieving 23.9, 23.5, 20.6, 16.2, and 3.7~Mb/s at positions~1--5, respectively.

The largest absolute advantage over LS/LMMSE occurs at NLoS position~4. RT-trained NRX attains 16.2~Mb/s, compared with 4.1~Mb/s for LS/LMMSE and 2.4~Mb/s for 3GPP-trained NRX. Using the unrounded means, these are approximately 4.0 and 6.8 times the respective baseline rates, with absolute increases of 12.2 and 13.8~Mb/s.

At LoS position~1, RT-trained NRX attains 23.9~Mb/s versus 15.5~Mb/s for LS/LMMSE and 9.5~Mb/s for 3GPP-trained NRX. The differences are smaller at positions~2 and~3. At position~5, the mean rate drops to 3.7~Mb/s for RT-trained NRX and 3.4~Mb/s for LS/LMMSE, while 3GPP-trained NRX records 0~Mb/s.

\subsection{Discussion}
This work is mainly applicable to SISO studies. Future work can extend it to MIMO scenarios and further optimize the neural network to further reduce inference latency.

\section{Conclusion}
A corridor digital twin built from a radar scan supplied ray-traced training channels for a 5G NR neural receiver deployed on DGX Spark and USRP B210 hardware. Under common radio settings, the receiver achieved higher mean uplink rates than LS/LMMSE and a 3GPP-trained neural receiver at all five test positions. At NLoS position~4 it reached 16.2~Mb/s, compared with 4.1 and 2.4~Mb/s for the two baselines. The testbed demonstrates how GPU-based neural detection can operate within the standard NR uplink processing chain.

\bibliographystyle{IEEEtran}
\bibliography{references}

@misc{cammerer2023,
  author = {Sebastian Cammerer and Fay{\c{c}}al A{\"i}t Aoudia and Jakob Hoydis and Andreas Oeldemann and Andreas Roessler and Timo Mayer and Alexander Keller},
  title = {A Neural Receiver for {5G NR} Multi-user {MIMO}},
  year = {2023},
  note = {arXiv:2312.02601},
  url = {https://arxiv.org/abs/2312.02601}
}

@misc{wiesmayr2024,
  author = {Reinhard Wiesmayr and Sebastian Cammerer and Fay{\c{c}}al A{\"i}t Aoudia and Jakob Hoydis and Jakub Zakrzewski and Alexander Keller},
  title = {Design of a Standard-Compliant Real-Time Neural Receiver for {5G NR}},
  year = {2024},
  note = {arXiv:2409.02912},
  url = {https://arxiv.org/abs/2409.02912}
}

@misc{baytekin2026,
  author = {Nuri Berke Baytekin and Reinhard Wiesmayr and Sebastian Cammerer and Chris Dick and Christoph Studer},
  title = {Site-Specific Finetuning of Neural Receivers with Real-World {5G NR} Measurements},
  year = {2026},
  note = {arXiv:2603.09644},
  url = {https://arxiv.org/abs/2603.09644}
}

@misc{luostari2026,
  author = {Riku Luostari and Dani Korpi and Harri Holma and Olav Tirkkonen},
  title = {Cross-Site Transfer and Spatial Forgetting in {OTA}-Trained Neural Receivers},
  year = {2026},
  note = {arXiv:2610.02577},
  url = {https://arxiv.org/abs/2610.02577}
}

@misc{sionna2022,
  author = {Jakob Hoydis and Sebastian Cammerer and Fay{\c{c}}al A{\"i}t Aoudia and Avinash Vem and Nikolaus Binder and Guillermo Marcus and Alexander Keller},
  title = {Sionna: An Open-Source Library for Next-Generation Physical Layer Research},
  year = {2022},
  note = {arXiv:2203.11854},
  url = {https://arxiv.org/abs/2203.11854}
}

@misc{sionnart2023,
  author = {Jakob Hoydis and Fay{\c{c}}al A{\"i}t Aoudia and Sebastian Cammerer and Merlin Nimier-David and Nikolaus Binder and Guillermo Marcus and Alexander Keller},
  title = {Sionna {RT}: Differentiable Ray Tracing for Radio Propagation Modeling},
  year = {2023},
  note = {arXiv:2303.11103},
  url = {https://arxiv.org/abs/2303.11103}
}

@techreport{3gpp38214,
  author = {{3GPP}},
  title = {{NR}; Physical Layer Procedures for Data},
  institution = {ETSI},
  number = {TS 138 214 V17.8.0},
  year = {2024}, month = feb,
  note = {3GPP TS 38.214, Release 17}
}

@misc{neuralrxsoftware,
  author = {Sebastian Cammerer and Reinhard Wiesmayr and Fay{\c{c}}al A{\"i}t Aoudia and Jakob Hoydis and Tommi Koivisto and Jakub Zakrzewski and Ruqing Xu and Pawel Morkisz and Chris Dick and Alexander Keller},
  title = {Real-time {5G NR} Multi-user {MIMO} Receivers},
  year = {2024},
  howpublished = {NVlabs software repository},
  url = {https://github.com/NVlabs/neural-rx}
}

@misc{sionna-rk,
  author = {Sebastian Cammerer and Guillermo Marcus and Tobias Zirr and Jakob Hoydis and Fay{\c{c}}al {Ait Aoudia} and Reinhard Wiesmayr and Lorenzo Maggi and Merlin Nimier-David and Alexander Keller},
  title = {{Sionna Research Kit}},
  year = {2026},
  howpublished = {NVIDIA software, version 1.3.3},
  url = {https://nvlabs.github.io/sionna/rk/index.html}
}
\end{document}